# Chemical vapor deposition synthesis of (GeTe)$_n$(Sb$_2$Te$_3$) gradient crystalline films as promising planar heterostructures

M. Zhezhu[1,*], A. Vasil'ev[1], M. Yapryntsev[2], E. Ghalumyan[3], D. A. Ghazaryan[4,5], H. Gharagulyan[1, 3, *]

[1]*A.B. Nalbandyan Institute of Chemical Physics NAS RA, Yerevan 0014, Armenia*
[2]*Belgorod State University, Belgorod 308000, Russia*
[3]*Institute of Physics, Yerevan State University, Yerevan 0025, Armenia*
[4]*Moscow Center for Advanced Studies, Kulakova str. 20, Moscow 123592, Russia*
[5]*Laboratory of Advanced Functional Materials, Yerevan State University, Yerevan 0025, Armenia*

[*]*The correspondence should be addressed to:*
*marina.zhezhu@ichph.sci.am and herminegharagulyan@ysu.am*

**Abstract**

**Phase-change materials of the (GeTe)$_n$(Sb$_2$Te$_3$) (GST) system are of high relevance in memory storage and energy conversion applications due to their fast-switching speed, high data retention, and tunable properties. Here, we report on a fast and efficient CVD-based method for the fabrication of crystalline GST films with variable Ge/Sb atomic content. In particular, the approach enables compositional control without changing the precursor, facilitating a gradient synthesis of Ge$_3$Sb$_2$Te$_6$, Ge$_2$Sb$_2$Te$_5$, and GeSb$_2$Te$_4$ phases in a single attempt. The analyses of their structural, optical, and electrical aspects highlight how compositional variation influences the film's properties. Our findings demonstrate a straightforward approach enabling the preparation of gradient crystalline GST films with tunable morphology and functionality. These gradient films can potentially provide in-plane multilevel and gradual switching thresholds for memory applications and altered refractive index and absorption for optical modulation and filtering applications.**

**Keywords:** phase-change materials, chemical vapor deposition, (GeTe)$_n$(Sb$_2$Te$_3$), gradient crystalline films.

**Abbreviations:** GST: (GeTe)$_n$(Sb$_2$Te$_3$); CVD: chemical vapor deposition.

## 1. Introduction

Phase-change materials based on the GeTe–Sb$_2$Te$_3$ pseudobinary system are best known for their use in optical data storage and nonvolatile memory devices [1, 2]. In recent years, their unique physical properties have also sparked a growing interest in emerging applications, such as thermoelectrics, sensors, neuromorphic computing, and infrared photonics [3-6]. (GeTe)$_n$(Sb$_2$Te$_3$) (GST) films are typically deposited using physical vapor deposition (PVD) methods, such as magnetron sputtering and pulsed laser deposition that produces amorphous films [7, 8]. Chemical vapor deposition (CVD) is a promising alternative to PVD methods for fabricating high-quality GST films [9, 10]. While PVD methods are widely used to produce homogeneous amorphous films, they often require high-vacuum systems and costly equipment. In addition, sputtered films are more prone to contamination from the sputtering gas and typically exhibit higher defect densities due to the high-energy impact of arriving atoms [11]. In contrast, CVD enables material transport and deposition at lower kinetic energies and under moderate vacuum conditions, thereby reducing equipment complexity and impurity incorporation. CVD also offers superior composition control, allowing for uniform and conformal film growth even on complex surfaces [12]. Although CVD processes may involve higher energy consumption, they provide better crystallinity and lower defect densities, which is particularly advantageous for the fabrication of multi-component and layered heterostructured films [13]. However, using metal-organic precursors in CVD processes raises significant safety concerns due to their toxicity and pyrophoric nature. Crystallization of the obtained amorphous films into either the metastable cubic phase or the thermodynamically stable hexagonal phase can subsequently be induced via Joule heating or by applying electrical, optical, or laser pulses [14, 15]. Additionally, superlattice-structured GST materials, such as $(GeTe)_n/(Sb_2Te_3)$, have recently attracted attention for advanced phase-change and thermoelectric applications due to their tailored electronic band structures, suppressed thermal conductivity, and enhanced switching performance [16-19]. Hexagonal phases of GST alloys demonstrate significant potential for thermoelectric applications owing to their combination of high electrical conductivity and low thermal conductivity [3]. Moreover, their favorable electronic structure and exceptional optical properties make them attractive for optoelectronic applications [20-22]. Among GST alloys, $Ge_3Sb_2Te_6$, $Ge_2Sb_2Te_5$, and $GeSb_2Te_4$—the most extensively studied composition—are located along the pseudobinary tie line between GeTe and $Sb_2Te_3$. All hexagonal GST compounds are narrow bandgap semiconductors that exhibit intrinsic $p$-type conduction. They are characterized by high electrical conductivity (~$10^5$ S/m) at room temperature [3]. Regarding optical properties, $Ge_2Sb_2Te_5$ is a promising broadband infrared absorber, attracting significant attention for optoelectronic applications [6]. Several studies highlighted the potential of this material as an efficient broadband absorber in the infrared region [22, 23]. Multilayer structures incorporating crystalline $Ge_2Sb_2Te_5$ demonstrated absorption levels as high as 93.6 % in the 1–1.6 $\mu$m wavelength range [6]. Furthermore, the inclusion of $Ge_2Sb_2Te_5$ significantly enhances the performance of plasmonic metasurface absorbers, achieving a maximum absorption of 99.56% in the 740–920 nm range for the hexagonal crystalline phase of the GST-coated layer [24]. Variation in their Ge/Sb/Te atomic ratios results in various structural characteristics and physical properties, providing a versatile platform for exploring structure-property relationships for future applications [3-6, 22-24]. While using crystalline gradient GST materials in photonics has not yet been widely explored, they hold significant potential for future applications. Particularly, a lateral compositional gradient in GST-based heterostructures offers a promising option for advanced photonic applications. It enables spatial tuning of optical properties, such as refractive index and absorption across the film, allowing for precise control of light propagation and interaction within a single device. This functionality can be exploited for optical modulation, where different regions of the film respond variably to external stimuli, like temperature and light, enabling dynamic modulation of transmission or reflection profiles. In spatial light control, the gradient can serve as a passive platform for beam shaping, focusing, and steering, with potential use in flat optics and metasurfaces [25-27]. Overall, the gradient GST design adds functional versatility and makes it suitable for compact, tunable, and highly integrated photonic devices.

This study aims to develop a simple method for the simultaneous CVD synthesis of GST films with tunable Ge/Sb compositions and to compare their structural, electrical, and optical properties for potential applications in electronic and optoelectronic devices. Designing gradient films of these materials with controlled electrooptical properties through compositional tuning allows them to be used in a single device. Specifically, we focus on the detailed analysis of the film's composition, microstructure, and morphology, aiming to reveal differences in elemental distribution across the deposited films and changes in morphology and grain size as a function of distance from the source. The gradient absorption properties, influenced by the film composition and crystallite size, open up opportunities for applying these films in near-infrared (NIR) optical sensors, where controlled signal attenuation is required, as well as in infrared (IR) shielding components and heat-absorbing photonic windows with spatially selective absorbance.

## 2. Experimental method

A $Ge_2Sb_2Te_5$ source crystal was grown beforehand to prepare the GST films. High-purity elemental precursors Ge (99.999%), Sb (99.999%), and Te (99.99%) were weighed in stoichiometric proportions corresponding to the nominal composition of Ge$_2$Sb$_2$Te$_5$. The reagents were sealed in an evacuated quartz ampoule, which was then gradually heated to 1000°C and held at this temperature for 2 h to ensure homogenization of the mixture and reaction between the elements. Subsequently, the ampoule was slowly cooled to allow the crystallization of the desired alloy. The resulting bulk alloy was cleaved using Scotch tape and characterized using X-ray diffraction (XRD), Raman spectroscopy, and scanning electron microscopy (SEM) methods.

Furthermore, this crystal, serving as a source material, was placed at the center of the heating zone inside a quartz tube with a diameter of 50 mm. Then, the polycrystalline Al$_2$O$_3$ substrates (1 × 1 cm) were positioned at varying distances (10–14 cm) from the source. During the deposition process, a constant vacuum of 10 Pa was maintained. The reaction tube was heated to 550°C over 30 minutes and held at this temperature for 30 minutes. The tube was rapidly removed from the heating zone to terminate the deposition. The GST films are labeled as follows: Film 1 corresponds to a source-to-substrate distance of 14 cm; Film 2 to 13 cm; Film 3 to 12 cm; Film 4 to 11 cm; and Film 5 to 10 cm. The obtained nanostructured films were characterized and analyzed using XRD analysis with grazing incidence (GIXRD, Rigaku Ultima IV diffractometer with Cu$K_\alpha$-radiation), SEM (Nova NanoSEM 450 microscope), atomic force microscopy and Raman spectroscopy (AFM-combined Raman spectrometer by LabRAM HR Evolution, HORIBA), optical spectroscopy (NIR Spectrometer 900-2500 nm, StellarNet, Inc.), and four-point probe electrical analysis (Veeco Instruments, Inc.).

## 3. Result and discussion

Fig. S1(a-c) presents the characterization of the synthesized source material. EDS analysis of the crystal fracture surface revealed the presence of 20.4 at.% Ge, 25.3 at.% Sb, and 54.3 at.% Te. The XRD diffractogram corresponds to the P$\bar{3}$m1 space group and confirms the formation of a single-phase Ge$_2$Sb$_2$Te$_5$ crystal. The Raman spectrum of the obtained crystal contains both sharp and broad spectral features. The sharp lines are attributed to phonon modes originating from Sb$_2$Te$_3$, which serves as the structural basis of GST, while the broad bands might be associated with Ge/Sb disorder [28]. Ge$_2$Sb$_2$Te$_5$, considered a quasibinary alloy of (GeTe)$_n$(Sb$_2$Te$_3$) with $n = 2$, represents an intermediate composition along the Sb$_2$Te$_3$-GeTe tie line. When used as a source material in the CVD process, it is expected to enable compositional gradient deposition of GST films due to preferential vapor transport and phase separation tendencies. As schematically illustrated in Fig. 1(a), the Ge$_2$Sb$_2$Te$_5$ crystal was placed in the hot zone, whereas the substrates were positioned in the cold zone at a distance of 10–14 cm. The Ge$_2$Sb$_2$Te$_5$ precursor was heated to 550°C since heating to 640°C and above can lead to the deposition of elemental (0-valent) Ge [29]. Initially, the composition, microstructure, and morphology of the films deposited on the substrates were analyzed (Fig. 1(b–f) and Fig. S2 (a–e)). EDS analysis demonstrates the differences in elemental distribution across the deposited films (Table 1). Here, Film 3, positioned at an intermediate distance, exhibits a Ge/Sb =1.

In contrast, Films 1 and 2 exhibit a higher Ge content relative to Sb, while Films 4 and 5 are enriched in Sb compared to Ge. Similar to the variation in elemental composition, morphology and grain size exhibit distinct changes as a function of distance from the source. Films 1–4 (see Fig. 1(b-e) and Fig. S2(a-d)) are mainly composed of plates growing perpendicular to the substrate, occasionally forming flower-like clusters and exhibiting plates with angular shapes. The plate size is characterized by the length (*l*) (Table 1), thickness (*d*), and *l/d* ratio (Table S2). The plate length gradually decreases with increasing source-to-substrate distance from 3.31 µm for Film 4 to 0.37 µm for Film 1. On the contrary, Film 5 displays signs of plate merging, with stacked partially fused or overlapped structures, indicating recrystallization after deposition. Fig. S2(a) clearly illustrates the small plates that form Film 1, while Fig. S2(e) reveals the overlapped, recrystallized plates.

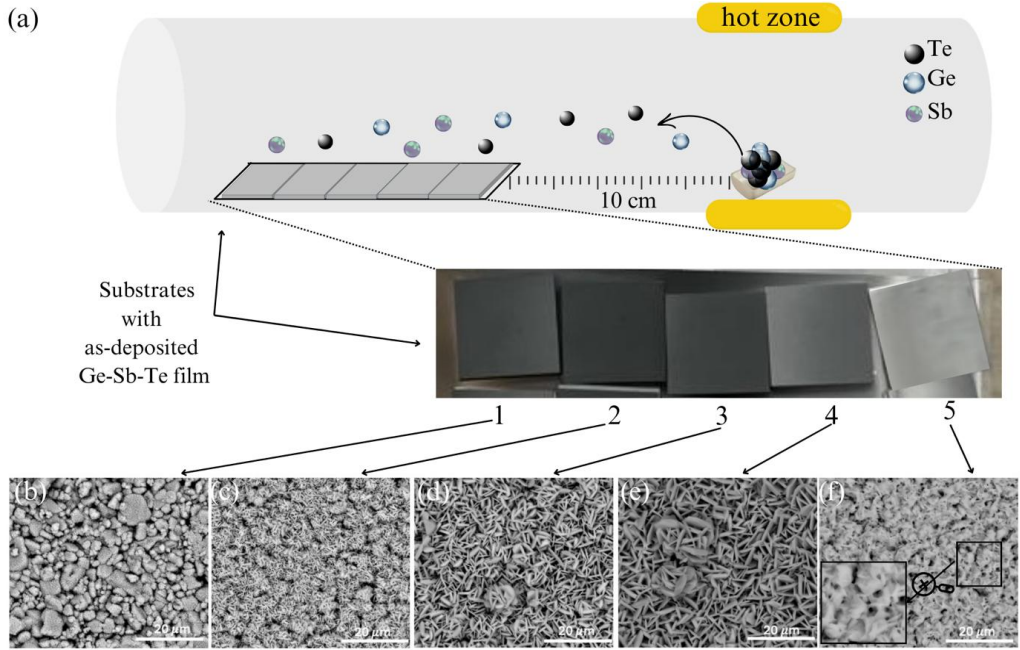

Fig. 1. Schematic illustration of the chemical vapor deposition process of GST films (a) along with the corresponding SEM images of the film's surface deposited on each substrate: Film 1 (b), Film 2 (c), Film 3 (d), Film 4 (e), and Film 5 (f), respectively. The zoomed-in area of Film 5 illustrates its recrystallization.

Additionally, $R_{ms}$ values were determined from AFM scans performed over an area of $10 \times 10$ μm$^2$ (Table 1). According to the obtained data, the films exhibit relatively high roughness, with the minimal $R_{ms}$ being 0.37 μm for Film 1. Thus, Film 1, positioned 14 cm from the source, shows the smallest plate size and lowest $R_{ms}$. Details of the morphological analysis are provided in the SI, and additional information on plate sizes is listed in Table S1.

Compared to crystalline GST films fabricated by magnetron sputtering [30], which typically exhibit a more homogeneous morphology and lower surface roughness after annealing of the as-deposited amorphous films, the CVD-grown films in our study demonstrate a distinctly different structure. These films consist of plate-like crystallites arranged into flower-like assemblies, resulting in a more textured surface with increased roughness. This morphological contrast arises from CVD enabling direct crystallization during deposition under varying temperature conditions, resulting in a more structurally diverse surface morphology [11].

A temperature gradient was established due to the source being located in the hot zone and the substrates positioned in the cold zone, with the gradient depending on the distance between them. To reliably assess the temperature distribution across the cold zone, a control (blank) experiment was conducted under the same thermal conditions as those used in the chemical deposition process. As a result, the deposition temperatures were determined to be approximately 195 °C for Film 1, 227 °C for Film 2, 286 °C for Film 3, 378 °C for Film 4, and 444 °C for Film 5, depending on the distance from the hot zone. The measurement scheme is shown in Fig. S3(a), and the obtained data are graphically presented in Fig. S3(b).

The deposition temperature is one of the most important parameters influencing the final grain structure of a vapor-deposited film for a given material system [11]. Specifically, according to classical nucleation theory, at low temperatures or in the presence of high diffusion barriers, adatoms - individual atoms that land on a surface during film deposition and can move or diffuse before becoming part of the solid film structure- tend to remain at their landing sites, resulting in amorphous or fine-grained polycrystalline film structures. As the temperature increases, enhanced surface diffusion allows adatoms to travel longer distances and coalesce into stable clusters, promoting the formation of larger grains [11]. Consistent with this, at elevated deposition temperatures, enhanced atomic diffusion decreases nucleation density, promoting the growth of larger grains and more continuous films. Conversely, lower temperatures limit atomic mobility, leading to higher nucleation density, smaller grain sizes, and a rougher surface morphology [31]. A correlation between increased deposition temperature, enlarged grain size, and surface morphology was also observed in GST films fabricated via the CVD method [32]. The thicknesses of the films were estimated from cross-sectional SEM images, and the obtained values range from 1.05 to 2.97 µm (Table 1).

The scaling of this deposition process would require the implementation of temperature-controlled zones of appropriate dimensions to ensure full coverage of the target area, the use of a larger-diameter quartz tube, and the ability to rapidly terminate the deposition process by shifting the quartz tube into a cold zone to stop material deposition immediately.

Table 1. Structural and morphological characteristics of the GST films.

| Sample | EDS analysis | | | Morphological analysis | | $t$ | XRD analysis | | | Film identification |
|---|---|---|---|---|---|---|---|---|---|---|
| | Ge, at. % | Sb, at. % | Te, at. % | $l$, µm | $R_{ms}$, µm | µm | Lattice parameters | | Space group | |
| | | | | | | | $a, b$ (Å) | $c$ (Å) | | |
| Film 1 | 28.0 | 16.6 | 55.4 | 0.37 | 0.37 | 1.16 | 4.2601 | 61.10 | $R\bar{3}m$ | $Ge_3Sb_2Te_6$ |
| Film 2 | 26.4 | 18.0 | 55.7 | 1.16 | 0.75 | 2.97 | 4.2104 | 61.718 | | |
| Film 3 | 22.4 | 21.9 | 55.7 | 2.11 | 0.57 | 1.27 | 4.2263 | 17.163 | $P\bar{3}m1$ | $Ge_2Sb_2Te_5$ |
| Film 4 | 16.4 | 28.4 | 55.4 | 3.31 | 0.59 | 1.05 | 4.2385 | 40.932 | $R\bar{3}m$ | $GeSb_2Te_4$ |
| Film 5 | 10.1 | 35.6 | 54.3 | Not determined | 0.20 | 0.70 | 4.2511 | 41.435 | | |

Based on the results of X-ray phase analysis (Fig. 2(a)), Films 1, 2, 4, and 5 belong to the hexagonal crystal system with the space group of $R\bar{3}m$, except for Film 3, which belongs to the $P\bar{3}m1$ space group. Phase identification was carried out considering the elemental composition investigated by the EDS detector of the electron microscope. Based on these data, Films 1 and 2 have the composition of $Ge_3Sb_2Te_6$, Film 3 corresponds to the $Ge_2Sb_2Te_5$ phase, and Films 4 and 5 are $GeSb_2Te_4$. Therefore, a source-to-substrate distance of 10–13 cm results in the deposition of GST with Ge/Sb ratio < 1, whereas a distance of 13-14 cm yields GST with a Ge/Sb ratio of 1, and a distance of 14-16 cm leads to the formation of GST with Ge/Sb ratio > 1. Moreover, X-ray phase analysis also reveals that in Films 2, 3, and 4, the peak corresponding to the (110) crystallographic plane is significantly enhanced compared to standard reference patterns, indicating a preferential grain orientation in these samples. Correlating these XRD results with the observed grain morphology suggests that the anisotropic plate-like crystallites are primarily oriented with their lateral facets exposed. Notably, numerous lateral facets can be associated with the (110) planes identified in the diffraction patterns. In contrast, for Film 5, the (110) peak is not enhanced, as the film underwent recrystallization, which diminished its anisotropy.

The space groups and corresponding lattice parameters are summarized in Table 1. The obtained lattice parameters are consistent with the values calculated in [33], where a dependence between the average interlayer distance and the cation-to-anion ratio for GeTe and $Sb_2Te_3$ binary compounds was established. Therefore, the observed variations in lattice parameters can be attributed to differences in the cation-to-anion ratio within the layered crystal structure of the GST compounds.

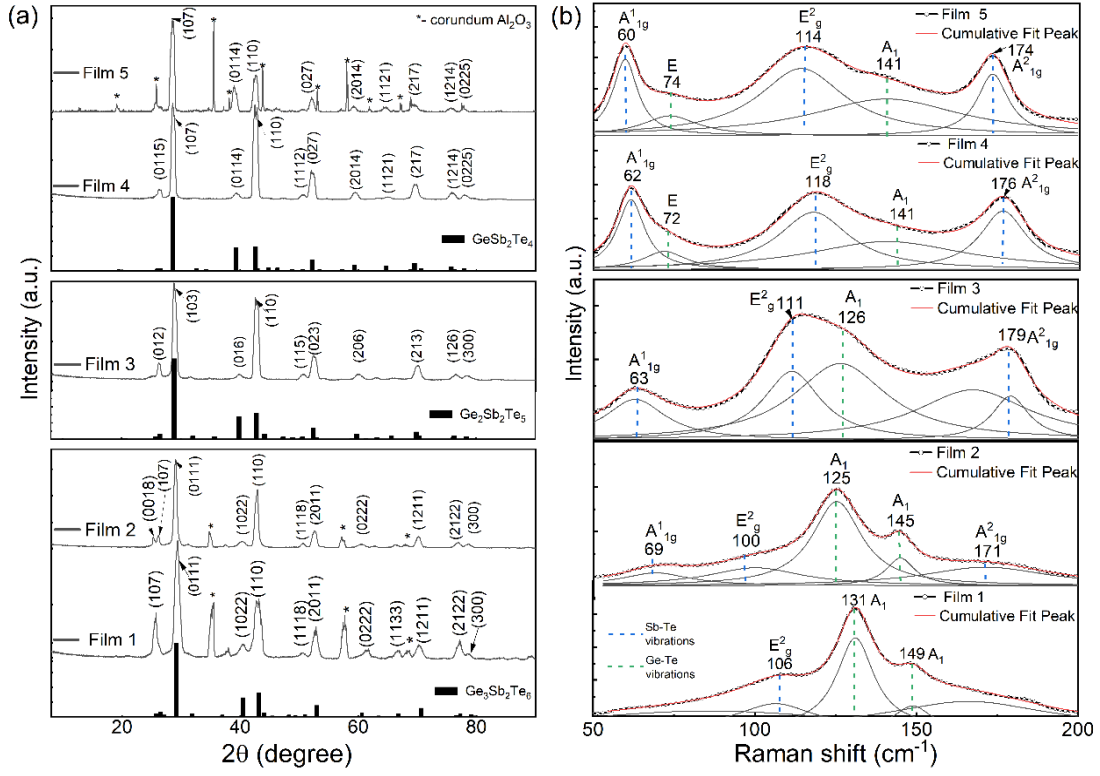

Fig. 2. Characterization spectra of the films obtained via GIXRD analysis (a) and Raman spectroscopy (b). The reference cards for the identified phases, along with the corresponding XRD diffractograms, are provided from the PDF-2 database: Card No. 01-077-6262 for $Ge_3Sb_2Te_6$, Card No. 01-073-7758 for $Ge_2Sb_2Te_5$, and Card No. 01-089-1064 for $GeSb_2Te_4$. The Raman spectra were analyzed using Lorentzian function fitting.

The hexagonal phase of GST adopts a layered tetradymite structure inherited from $Sb_2Te_3$. The stacking arrangement is influenced by the ratio of GeTe insertions. For $(GeTe)_n(Sb_2Te_3)$ ($n \geq 1$), the structure consists of identical $Sb_2Te_3$ slabs with $n$-GeTe insertions forming layers stacked along the $c$-axis via van der Waals gaps. Each slab contains ($2n + 5$) layers, as each GeTe insertion introduces two additional layers [34]. The $Ge_3Sb_2Te_6$, $Ge_2Sb_2Te_5$, and $GeSb_2Te_4$ structures identified in the films represent the lowest-energy configurations for the examined Ge/Sb compositions, aligning with previous reports [35].

Although GST films have been extensively studied, a comprehensive Raman characterization of their crystalline phases remains limited. In the current study, we present a Raman analysis of crystalline GST films with varying structural features (Fig. 2(b)). In particular, Raman spectra of the crystalline GST films reveal characteristic vibrational modes associated with both Ge-Te and Sb-Te bonds, enabling the identification of structural differences across the compositional gradient. The $Ge_3Sb_2Te_6$, $Ge_2Sb_2Te_5$, and $GeSb_2Te_4$ structures corresponding to Ge/Sb > 1, = 1, and < 1, respectively, can be described in terms of $n$-GeTe units ($n$ = 3, 2, and 1) within the $(GeTe)_n(Sb_2Te_3)$ framework. The Raman spectrum of $Ge_3Sb_2Te_6$ (Films 1 and 2) is characterized by $A_1$ modes corresponding to corner-sharing and edge-sharing $GeTe_{4-n}Ge_n$ ($n$ = 0 - 3) tetrahedra [14, 36], observed at ~ 131 cm$^{-1}$ and ~ 149 cm$^{-1}$ for Film 1 and at 125 cm$^{-1}$ and 145 cm$^{-1}$ for Film 2. These features reflect the dominance of Ge–Te vibrational modes in $Ge_3Sb_2Te_6$, which contains three GeTe units per $Sb_2Te_3$ unit. In $Ge_2Sb_2Te_5$ (Film 3), both Ge–Te and Sb–Te vibrational modes are observed. The presence of the $A_1$ corner-sharing $GeTe_{4-n}Ge_n$ ($n$ = 0 - 3) tetrahedra mode at 126 cm$^{-1}$, along with modes $A^1_{1g}$ at 63 cm$^{-1}$, $E^2_g$ at 111 cm$^{-1}$, and $A^2_{1g}$ at 179 cm$^{-1}$ of Sb$_m$Te$_3$ ($m$ = 1, 2) [14, 36], indicates a balanced contribution from both Ge-Te and Sb-Te subsystems. GeSb$_2$Te$_4$ (Films 4 and 5) exhibits dominant Sb–Te vibrational modes. In Film 4, $A^1_{1g}$ appears at 62 cm$^{-1}$, $E^2_g$ at 118 cm$^{-1}$, and $A^2_{1g}$ at 176 cm$^{-1}$; in Film 5, each of these modes redshifts by approximately 2 cm$^{-1}$, suggesting a more ordered Sb$_2$Te$_3$-like lattice. Also, Ge-Te vibrational features are presented, with the $E$ and $A_1$ modes appearing at 72 cm$^{-1}$ and 141 cm$^{-1}$ in Film 4 and at 74 cm$^{-1}$ and 141 cm$^{-1}$ in Film 5. Their reduced intensity further reflects the prevailing Sb$_2$Te$_3$-like structural character. Variations in the Raman modes correlate with the relative GeTe and Sb$_2$Te$_3$ content, reflecting structural gradient across the films.

Based on these data, the deposition process led to the gradual formation of GST films. To explain this process, the vapor pressures of the binary compounds GeTe and Sb$_2$Te$_3$, which constitute GST alloys, were analyzed. A review of the literature revealed that at a temperature of 550°C, the vapor pressure of GeTe is three times higher than that of Sb$_2$Te$_3$ (8.9 Pa [37] against 2.9 Pa [38]), facilitating the transport of germanium telluride over a greater distance from the source and enriches the more distant substrates with GeTe. Additionally, the temperature gradient, characterized by a decrease in temperature with increasing source-to-substrate distance, likely leads to reduced molecular kinetic energy, thereby contributing to the variation in film thicknesses during the deposition process [39, 40]. Consequently, a small local chemical gradient results in the formation of in-plane heterostructure crystalline films composed of the most stable GST structures, as shown in the compositional diagram (Fig. 3(a)). The obtained crystalline films lie at the pseudobinary line between GeTe and Sb$_2$Te$_3$.

Further, the electrical and optical properties of the above-mentioned films were studied (Fig. 3(b, c)). As is known, the electrical resistivity $\rho = \frac{1}{nq\mu}$ where $n$ is the carrier concentration, $q$ is the carrier charge, and $\mu$ is the carrier mobility [41]. Despite a similar carrier concentration at room temperature reported in [3, 42] for Ge$_3$Sb$_2$Te$_6$, Ge$_2$Sb$_2$Te$_5$, and GeSb$_2$Te$_4$, the synthesized films exhibit notable differences in resistivity, which we believe correlate with crystallite size rather than the composition of the film. Ge$_3$Sb$_2$Te$_6$ films with the smallest plates show the highest resistivity (~17.04×10$^{-5}$ Ω·m), while GeSb$_2$Te$_4$ films with larger plates exhibit lower values (~3.50×10$^{-5}$ Ω·m). This trend reflects reduced carrier scattering at grain boundaries with increasing plate size, thereby enhancing carrier mobility. So, the observed differences in resistivity among the films are primarily attributed to variations in crystallite size, with larger plates leading to reduced grain boundary scattering and enhanced carrier mobility.

The absorption spectra of proposed crystalline GST compositions were measured at normal incidence of light in the infrared range at 950–1600 nm. Remarkably, the evaluation of the absorbance normalized to a film thickness ($A/t$) for Ge$_3$Sb$_2$Te$_6$, Ge$_2$Sb$_2$Te$_5$, and GeSb$_2$Te$_4$ films revealed values ranging from ~ 0.5 to ~ 0.98 $\mu$m$^{-1}$. The films exhibit broadband absorption behavior across the NIR region. The Ge$_2$Sb$_2$Te$_5$ film had the highest absorption (0.88 ÷ 0.98 $\mu$m$^{-1}$), whereas Ge$_3$Sb$_2$Te$_6$, particularly the sample with smaller crystallites, had the lowest absorption (0.34 ÷ 0.25 $\mu$m$^{-1}$). The absorption for GeSb$_2$Te$_4$ films stayed between Ge$_3$Sb$_2$Te$_6$ and Ge$_2$Sb$_2$Te$_5$, varying with surface morphology. Film 5, characterized by partially fused plates, exhibits absorption in the range 0.55 to 0.5 $\mu$m$^{-1}$, whereas Film 4 demonstrates higher absorption values, ranging from 0.64 to 0.71 $\mu$m$^{-1}$. The reflectance of the films normalized to a film thickness was also measured (Fig. S4). These variations are attributed to compositional differences and film morphology: An increase in crystallite size improves light absorption by minimizing scattering at grain boundaries. Among the above-mentioned structures, Ge$_2$Sb$_2$Te$_5$ was expected to exhibit high absorbance. Furthermore, additional absorption measurements for the above-mentioned films on the KBr substrate were performed in the 6–20 $\mu$m range, where the films showed strong absorption (Fig. S5).

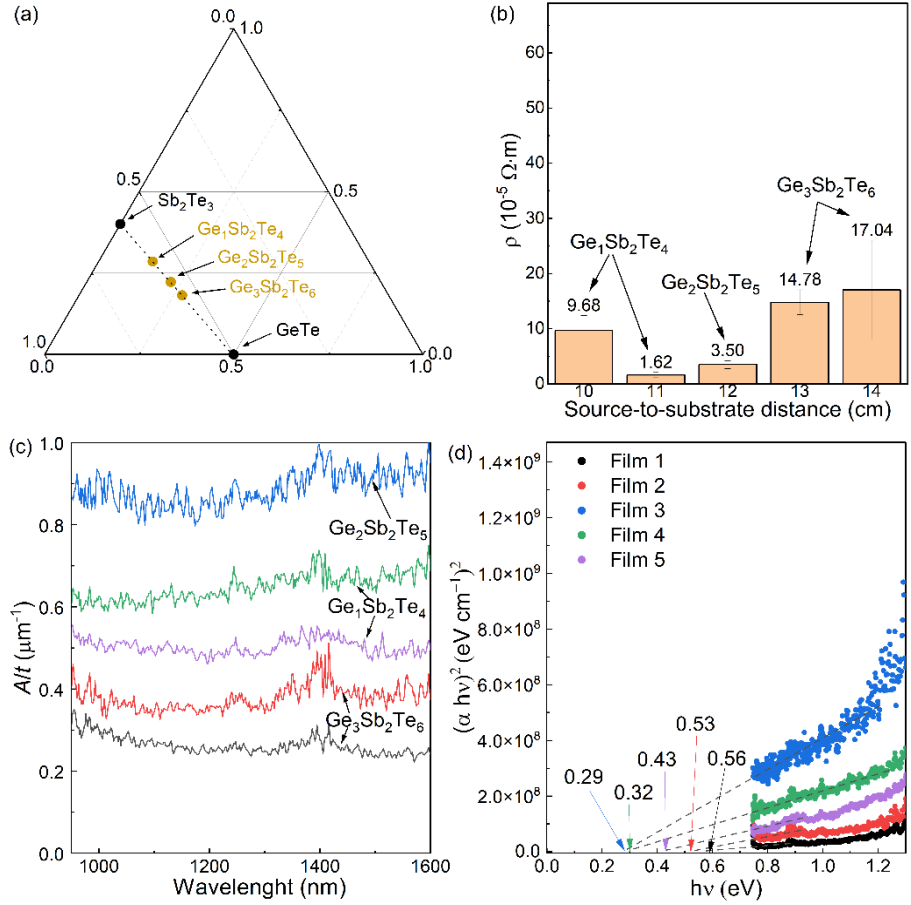

Fig. 3. (a) Compositional diagram of Ge–Sb–Te system indicating the positions of the synthesized films along the Sb$_2$Te$_3$–GeTe tie line. (b) The dependence of specific electrical resistivity, including relative errors on the source-to-substrate distance. The sheet resistance was measured 10 times per film. (c) The thickness-normalized absorption spectra measured in the range of 950–1600 nm. All measurements were conducted at room temperature. (d) Estimation of the optical band gap of the synthesized GST films using Tauc plot analysis based on their absorbance spectra.

The optical bandgaps of the crystalline GST films were extracted from absorption spectra using Tauc plot analysis. The studies employed a Tauc exponent of $r = 1/2$ corresponding to direct allowed electronic transitions in crystalline semiconductors [43, 44]. Band structure calculations reveal minor differences between the indirect and direct bandgaps in the crystalline phases of the (GeTe)$_n$(Sb$_2$Te$_3$) system [45]. Fig. 3(d) indicates that the minimum optical band gap of 0.29 eV was obtained for the Ge$_2$Sb$_2$Te$_5$ film. GeSb$_2$Te$_4$-based films (Films 4 and 5) exhibited optical band gaps of 0.32 eV and 0.43 eV, respectively. Finally, the Ge$_3$Sb$_2$Te$_6$ films (Films 1 and 2) demonstrated wider band gaps of 0.56 eV and 0.53 eV, respectively. Previous reports revealed that hexagonal Ge$_2$Sb$_2$Te$_5$ and many other compositions within the (GeTe)$_n$(Sb$_2$Te$_3$) system behave as degenerate p-type semiconductors with the Fermi level located right above the valence band [46]. So, the optical bandgap is typically smaller, with values of 0.24 to 0.6 eV for the crystalline phase [3, 47, 48]. Although amorphous GST materials usually exhibit a direct bandgap (0.7–0.8 eV) and crystalline GST show indirect bandgaps (0.4-0.6 eV) [32], the strain at the grain boundaries may contribute to the modification of the electronic band structure leading to a shift toward direct bandgap behavior similar to that in [49, 50].

Another key aspect of these materials involves the texture-dependent anisotropic properties. Particularly, it is well established that two-dimensional materials with layered crystal structures exhibit pronounced anisotropic grain growth primarily due to the disparity in bonding within and between the layers: covalent in-plane and van der Waals out-of-plane, respectively [51]. In the case of our hexagonal GST films, the preferential orientation corresponds to the lateral (110) plane being parallel to the substrate surface. Given the layered nature of the crystal structure, charge carrier transport is expected to proceed more efficiently along the covalently bonded *ab* plane than along the *c*-axis owing to reduced carrier scattering in the in-plane direction. This behavior is in agreement with numerous reports on the anisotropic properties of two-dimensional chalcogenide materials [51-53]. Regarding the effect of texture on optical properties, it has been shown that preferential orientation can significantly impact the reflectance. Specifically, literature [54] reports an increase in optical reflectance associated with a texture transition from the (11*l*) plane to the (00*c*) plane in Sn-doped crystalline GST films. Thus, in the case of our films, the optical reflectance associated with the (11*l*) orientation is expected to be reduced compared to that from the *ab* planes.

## 4. Conclusion

A fast and efficient approach was developed for the CVD synthesis of gradient crystalline GST films with varying Ge/Sb compositions. By simply adjusting the source-to-substrate distance without changing the precursor, the films with compositions of $Ge_3Sb_2Te_6$, $Ge_2Sb_2Te_5$, and $GeSb_2Te_4$ were simultaneously deposited onto substrates potentially allowing controllable formation of planar GST gradient heterostructures. The films' structural properties, electrical resistivity, and optical absorbance were analyzed to evaluate their potential applications in electronic and optoelectronic devices. In particular, an in-depth analysis of the composition, microstructure, and morphology of the GST films uncovered variations in elemental distribution, morphology, and grain sizes relative to their distance from the source crystal adjusted during growth. We believe that using a large-area substrate during the growth of GST films may allow for the potential fabrication of gradient IR filters—exhibiting high central absorption and lower absorption at the edges—which may be beneficial for photonic devices, such as optical modulators, beam shaping elements, and photonic circuits with built-in spatial control functionalities or sensors requiring spatially resolved control of signal intensity.

# Supplementary Information for
# Chemical vapor deposition synthesis of (GeTe)$_n$(Sb$_2$Te$_3$) gradient crystalline films as promising planar heterostructures

M. Zhezhu[1,*], A. Vasil'ev[1], M. Yapryntsev[2], E. Ghalumyan[3], D. A. Ghazaryan[4,5], H. Gharagulyan[1, 3, *]

[1] A.B. Nalbandyan Institute of Chemical Physics NAS RA, Yerevan 0014, Armenia
[2] Belgorod State University, Belgorod 308000, Russia
[3] Institute of Physics, Yerevan State University, Yerevan 0025, Armenia
[4] Moscow Center for Advanced Studies, Kulakova str. 20, Moscow 123592, Russia
[5] Laboratory of Advanced Functional Materials, Yerevan State University, Yerevan 0025, Armenia

[*]The correspondence should be addressed to:
marina.zhezhu@ichph.sci.am and herminegharagulyan@ysu.am

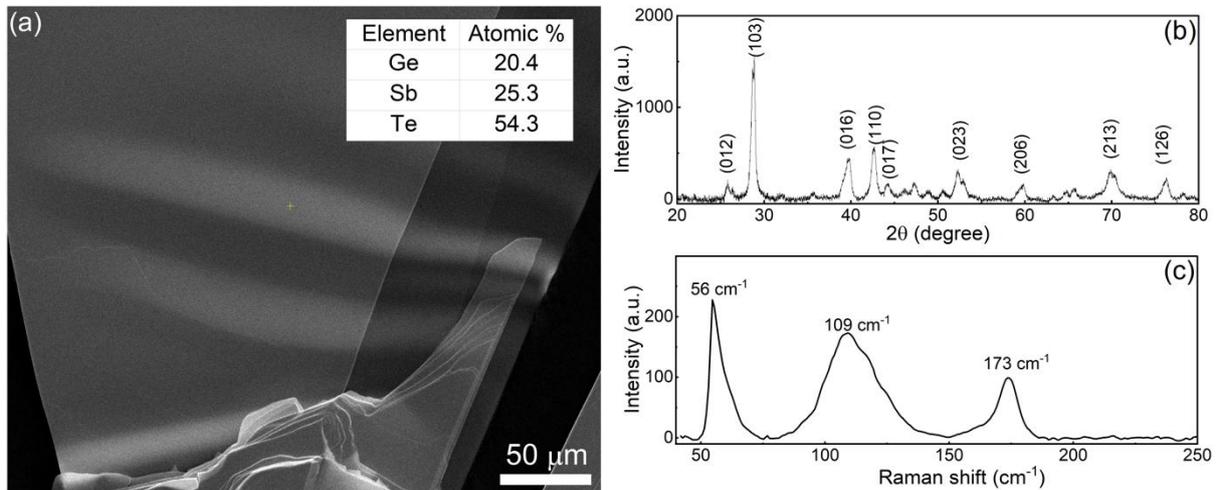

Fig. S1. Characterization of the synthesized target includes SEM analysis of the fracture surface with (a) EDS, (b) XRD analysis, and (c) Raman spectroscopy.

*Details of morphological analysis*

SEM images revealed variations in grain structure, while AFM measurements provided insights into surface roughness and topographical features. To obtain a statistically reliable *l/d* ratio, the length and thickness of over 150 individual plates were measured from SEM images. The resulting distributions for both parameters are well described by a unimodal lognormal function, and the corresponding values are presented in Table S1. Additionally, $R_{ms}$ values were determined from 10 μm × 10 μm AFM scans.

Table S1. Plate sizes represented by *d* and *l/d* ratio of Films 1–4.

| Film number | *d*, nm | *l/d* |
|---|---|---|
| 1 | 143 | 2.6 |
| 2 | 224 | 5.2 |
| 3 | 277 | 7.6 |
| 4 | 543 | 6.1 |

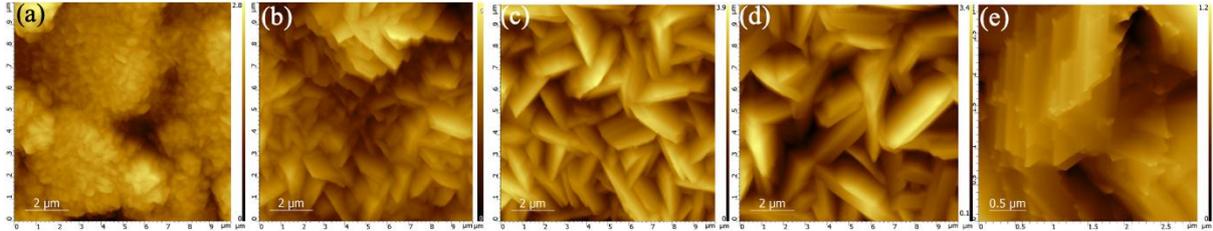

Fig. S2. The AFM images of the film surfaces deposited on each substrate: (a) Film 1, (b) Film 2, (c) Film 3, (d) Film 4, and (e) Film 5.

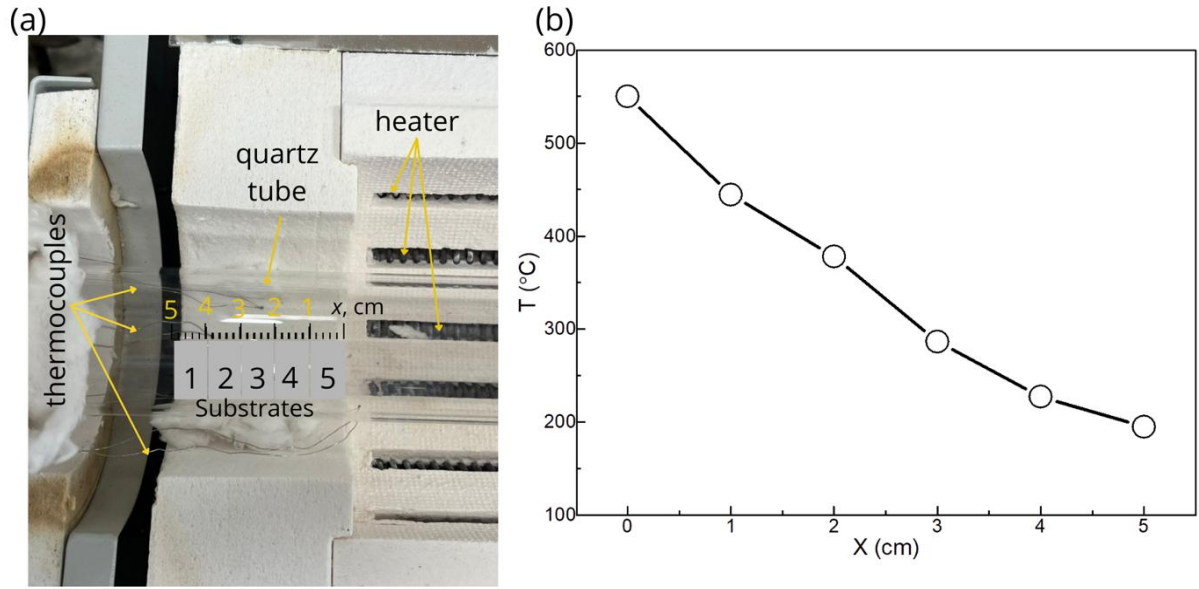

Fig. S3. Control experiment schematic for measuring temperature distribution in the cold zone. (b) Temperature gradient across the cold zone as a function of distance from the hot zone. The segment from 0 to 1 cm corresponds to the position of Film 5; 1–2 cm to Film 4; 2–3 cm to Film 3; 3–4 cm to Film 2; and 4–5 cm to Film 1.

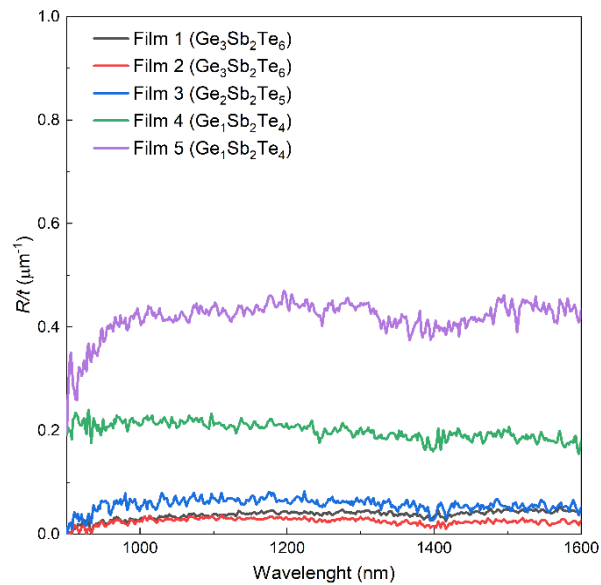

Fig. S4. The thickness-normalized reflectance spectra measured in the range of 950–1600 nm at room temperature.

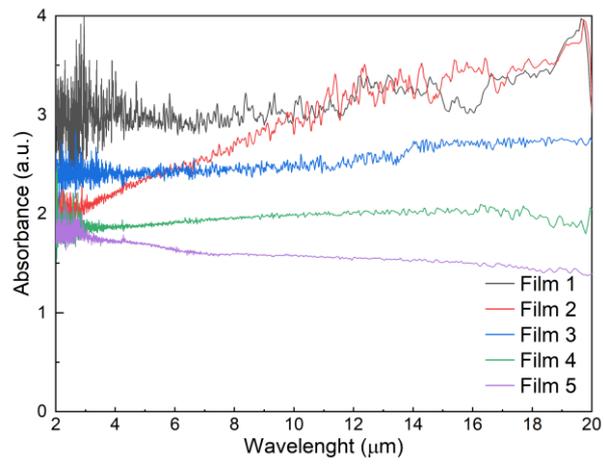

Fig. S5. Absorbance spectra measured on a KBr substrate in the 2–20 μm wavelength range at room temperature.